# Laser pulse compression by a density gradient plasma for exawatt to zettawatt lasers

Min Sup Hur[1,*], Bernhard Ersfeld[3], Hyojeong Lee[2], Hyunseok Lee[1], Kyungmin Rho[2], Yunkyu Lee[1], Hyung Seon Song[1], Samuel Yoffe[3], Dino A. Jaroszynski[3,†], and Hyyong Suk[2,‡]

[1]Department of Physics, UNIST, 50 UNIST-gil, Ulju-gun, Ulsan, 689-798, Republic of Korea
[2]Department of Physics and Photon Science, Gwangju Institute of Science and Technology, Gwangju, Republic of Korea
[3]Department of Physics, Scottish Universities Physics Alliance and University of Strathclyde, Glasgow G4 0NG, UK

[*]Electronic address: mshur@unist.ac.kr
[†]Electronic address: d.a.jaroszynski@strath.ac.uk
[‡]Electronic address: hysuk@gist.ac.kr

**Abstract**

We propose a new method for laser pulse compression that uses the spatially varying dispersion of a plasma plume with a density gradient. This novel scheme can be used to compress ultrahigh power lasers. A long, negatively frequency-chirped, laser pulse reflects off the plasma ramp of an over-dense plasma. As the density increases longitudinally the high frequency photons at the leading part of the laser pulse propagates more deeply than low frequency photons, the pulse is compressed in a similar way to compression off a chirped mirror. Proof-of-principle simulations, using a one-dimensional (1-D) particle-in-cell (PIC) simulation code demonstrates the compression of 2.35 ps laser pulse to 10.3 fs, with a compression ratio of 225. As plasmas is robust and resistant to high intensities, unlike gratings in a chirped-pulse amplification (CPA) technique [1], the method could be used as a compressor to reach exawatt or zettawatt peak power lasers.

# I. INTRODUCTION

The invention of the chirped-pulse-amplification, CPA, technique [1] has made possible the development of ultra-short pulse, multi-petawatt (PW) laser systems. The PW-pulses are useful for plasma-based compact accelerators that can produce multi-GeV electron beams [2] and hundreds of MeV ion beams [3], which otherwise are only available at large scale accelerating facilities. Exawatt (EW, 1 EW= $10^{18}$ W) laser pulses provide tools to experimentally study various outstanding problems of modern theoretical physics, such as the pair production [4], photon-photon scattering [5], radiation reaction [6] and lab-astrophysics for the early universe [7]. Zettawatt (ZW, 1 ZW = $10^{21}$ W) laser pulse would exceed the Schwinger limit [6] and address the paradox of information loss in blackhole [9].

The ZW lasers are not likely to be realised using current CPA technique because manufacturing of large-size compression gratings is almost at its technological limit. Even for PW-level pulse compression, metre-scale gratings are required to keep the pulse intensity below the surface damage threshold. From simple extrapolation, the diameter of gratings for EW or ZW lasers would be hundreds of metres in width. This is practically impossible and fundamentally different approaches to compression should be taken. As material damage is the major obstacle on the path toward post-EW powers, a plasma-based compression scheme is very appealing. Because plasma is an already broken-down state of matter, it is fundamentally damage-resistant to high electric field of intense laser pulses. The maximum energy density sustainable by plasma is several orders of magnitude higher (>$10^{16}$ W/cm$^2$) than that of solid state media (~ $10^{13}$ W/cm$^2$) [10,11]. This means that metre-scale gratings can potentially be replaced by millimetre-size plasma if a method can be found to make the plasma dispersive. One promising plasma-based method for pulse compression is Raman (or Compton) amplification [12-19]. In this scheme, a long pump laser pulse is backward-scattered and compressed by an electron plasma wave (or bunches) that is self-generated under the action of the ponderomotive force of the beat wave associated with the pump and counter-propagating seed pulse (forming the compressed pulse). Compression of 25 ps-long, 4 TW (>100 J) pulse into 25 fs, 2 PW (~50 J) (efficiency of 35 %) has been obtained virtually from PIC (particle-in-cell) simulations [17]. Experimentally, compression of Joule-level pulses (~50 J) was first demonstrated with 10% efficiency [19]. A slower, but more robust amplification by Brillouin scattering, has also been studied [20]. Generation of plasma gratings by beating two counter-propagating laser pulses as a compressor has been suggested [21-24]. Compression using plasma gratings by a small factor has been shown in PIC simulations [23,24], but further noteworthy results have yet to come.

Here we introduce a fundamentally different approach. Our method is much simpler and most of all, considerably more efficient than other proposed plasma-based schemes. We demonstrate that a long chirped pulse can be compressed several hundred times to intensities of $I \sim 10^{17}$ W/cm$^2$, with almost no energy loss, using a millimetre-size plasma, which is orders-of-magnitude smaller than gratings of conventional CPA systems. By further optimization it is expected that zettawatt powers should be obtainable from a compact system.

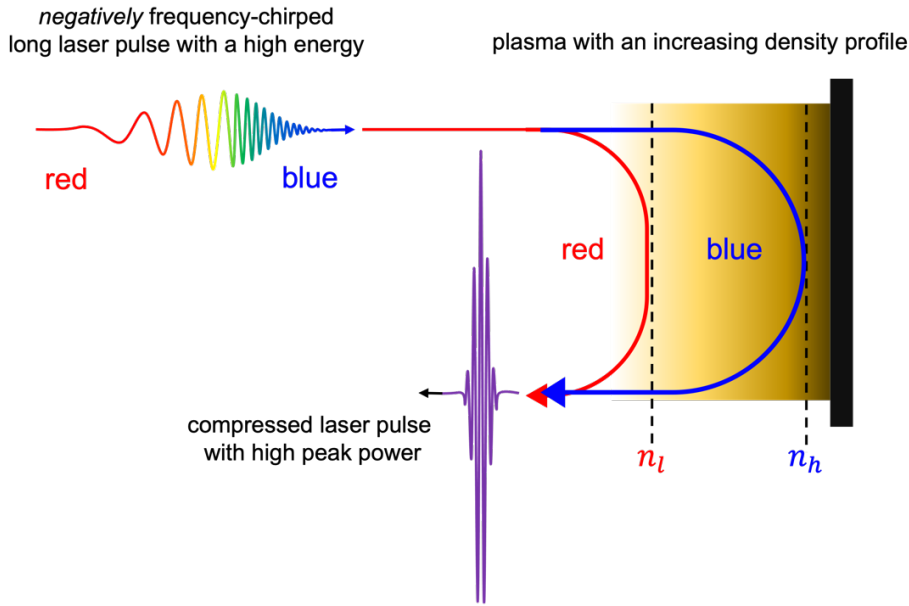

FIG. 1: The laser pulse compressor based on a density gradient plasma, where a long, high energy, negatively frequency-chirped laser pulse reflects for an over-dense plasma that has an increasing density ramp that has spatially dependent dispersion. High frequency components at the front of the pulse and low frequency components at the rear part of the pulse reflect at different positions, leading to the compression of the laser pulse.

## II. BASIC IDEA

The plasma-based schemes of laser pulse compression described in the introduction utilise plasma that has a layered density profile to scatter a pump pulse. However, plasma is intrinsically a dispersive medium, even without periodic structures. Here, we propose a new and more direct way of exploiting the dispersive property of plasma for the pulse compression. The dispersion relation of an electromagnetic wave of frequency $\omega$ in a plasma is given by $\omega^2 = \omega_p^2 + c^2 k^2$, where the plasma frequency is $\omega_p = (n_0 e^2 / m \epsilon_0)^{1/2}$ and the group velocity of photons depends on frequency. Most importantly, a photon is reflected at the point where its frequency is identical to $\omega_p$; photons with higher frequency are reflected at higher density. Therefore, when a negatively frequency-chirped laser pulse is incident on a plasma with an increasing density, higher-frequency photons have longer round-trip paths than lower-frequency ones, resulting in compression of the laser pulse (Fig. 1).

This process is similar but has differences, compared with laser pulse compression process in a CPA system. In the CPA technique, a long duration (100s ps to ns), chirped (mostly positive) laser pulse and ruled gratings are used to provide the correct dispersion. In our idea, a much shorter duration, negatively-frequency-chirped laser pulse and a damage-free plasma, which can sustain a much higher intensity laser pulse, are employed. Examples of one-dimensional (1D) PIC (particle-in-cell) simulations show that a 2.35 ps (FWHM), frequency-chirped ($\Delta\omega/\omega = 0.13$) Gaussian laser pulse with a peak intensity of $I_0 = 5.44 \times 10^{14}$ W/cm² is increased to $I = 9.3 \times 10^{16}$ W/cm², a factor of 170 times larger, in a 2.5 mm-long plasma that has a quadratically increasing density profile. Previous studies [10,11] have shown that the damage threshold of conventional gratings in CPA is about $10^{13}$ W/cm². Therefore, the intensity of the compressed laser pulse in our simulation is at least 3 orders of magnitude higher than that of a conventional CPA compressors. This implies that a density gradient plasma with a diameter of only 10 cm can be used to reach 7.3 EW laser power, which is unprecedented. From our other simulation, increasing the intensity by 225 times, corresponding to 2.3 EW, for the same assumed diameter, has been obtained. The new method proposed here has a potential to open up a new possibility for exawatt-and-above high power lasers in the future.

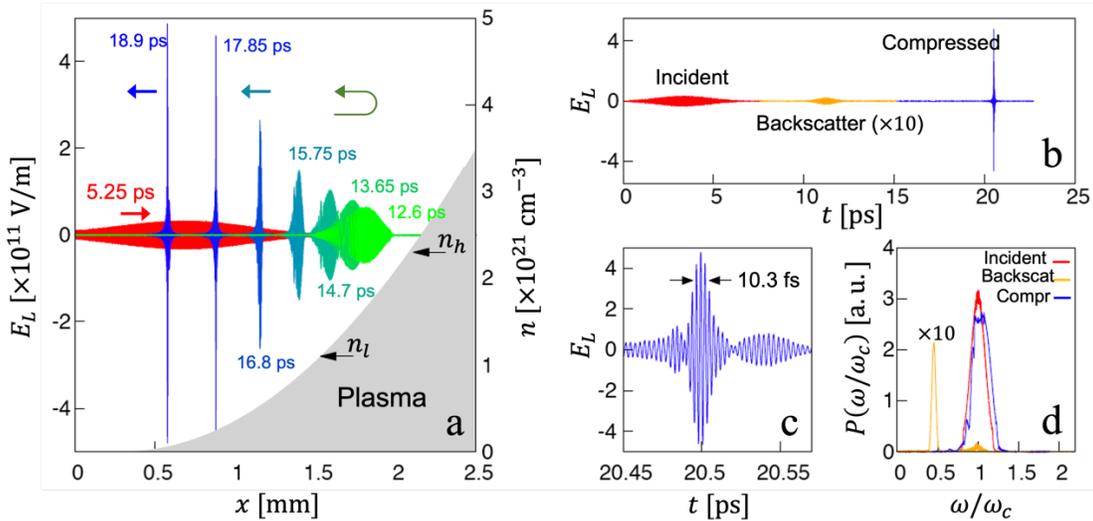

FIG. 2: Pulse compression in a gradient plasma from a 1D PIC simulation. **a.** Incident pulse (red) and stroboscopic view of the reflected pulse (green to blue). On the plasma density profile (grey shade), $n_h$ and $n_l$ are the critical densities for $\omega_h$ (highest frequency at the front tip of the pulse) and $\omega_l$ at the rear tip, respectively. **b.** Electric field passing through a virtual probe located at $x = 0.1$ mm as a function of time. The data is split into the incident (red), the compressed (blue) and the Raman-backscattered (RBS, orange). The RBS is 10-times magnified. **c.** The compressed pulse in time. **d.** Power spectra of the incident (red), RBS (orange, 10-times magnified) and the compressed (blue).

## III. SIMULATION RESULTS

Figure 2 presents 1D PIC simulation results showing a factor of 225 gain in intensity due to pulse compression. The incident pulse (red in Fig. 2a) of duration of 2.35 ps FWHM (full-width-at-half-maximum), is compressed to 10.3 fs FWHM (blue). The peak amplitude of the electric field grows by a factor of 15, from $3.22 \times 10^{10}$ V/m ($I_0 = 1.37 \times 10^{14}$ W/cm$^2$) to $4.79 \times 10^{11}$ V/m ($I = 3.05 \times 10^{16}$ W/cm$^2$). The incident laser pulse is longitudinally Gaussian $\sim \exp[-t^2/\sigma^2]$, with a cut-off at $1.5\sigma$. The pulse duration is set to $\sigma = 2$ ps (2.35 ps FWHM). The angular frequency is negatively chirped starting from $\omega_h = 2.73 \times 10^{15}$ rad/s ($\lambda_h = 690$ nm), sweeping linearly to $\omega_l = 1.86 \times 10^{15}$ rad/s ($\lambda_l = 1010$ nm). The corresponding relative bandwidth $\Delta\omega_{FHWM}/\omega_c = 0.13$ indicates a Fourier limited pulse of 9.8 fs, which would be the minimum pulse duration possible by compression. A cold ($T_e = 0$) plasma is loaded with quadratically increasing density profile given by the function $n(x) = n_h(x - x_0)^2/L^2$, where we set $L = 1.81$ mm, $x_0 = 0.3$ mm (plasma-vacuum boundary) and $n_h = 2.35 \times 10^{21}$ cm$^{-3}$, where $n_h$ is the critical density for $\omega_h$. The density ramp is extended to $1.5n_h$ to minimize leakage of pulse energy beyond the critical point through the skin depth. Simulation particles are loaded by the cumulative distribution technique [25] to minimize artificial noise in the density profile. Note that, for a quadratic density profile, the round-trip time of photons is a linear function of frequency (i.e. $\int_0^{l_c} v_g^{-1} dx \propto \omega$, where $l_c$ is the critical point), which is compatible with compression of linearly-chirped pulses.

In the simulation, a virtual probe is placed at $x = 0.1$ mm (vacuum side) to "detect" the electric field as a function of time (Fig. 2b) and evaluate the frequency spectra of both the incident and reflected pulses (Fig. 2d). The probed data are split into three sections; the incident pulse (red in Fig. 2b), the compressed pulse (blue) and the pre-backscattered portion in the middle (orange). The compressed pulse (Fig. 2c) has 99.2 % of the incident energy. The original spectrum of the incident pulse is well preserved in the compressed pulse (Fig. 2d). A small portion of the incident energy is lost via Raman backscatter (Fig. 2b,d). Very small energy loss during the compression is one of the advantages of our method over other plasma-based compression schemes. As will be shown later, under more realistic conditions, a fraction of the original pulse is lost due to various mechanisms, but many of these can be kept under control.

For a warm plasma, thermal noise can affect compression in diverse ways. In 1D, energy loss by Raman backscatter (RBS) is the most deleterious effect. The thermal noise can trigger RBS much earlier than in cold plasma. Figure 3 shows the simulation result from a thermal plasma with a temperature of 10 eV, with all other parameters kept the same as in Fig. 2. A significant portion of the rear part of the pulse is depleted by RBS (Fig. 3a,b). Note that RBS is triggered at the quarter-critical density, i.e. when $n_{1/4} = 0.25 n_c$ ($n_c$ is the critical density for the central frequency $\omega_c$), where the growth rate of RBS is the maximum. The peak amplitude ($\sim 3.7 \times 10^{11}$ V/m) after the compression is slightly reduced from the cold case (Fig. 3c). The total compressed energy is 75 % of the incident pulse energy. Scattered energy by RBS is 16 % of the incident pulse. The remaining 10 % is spread out by thermal fluctuation. Figure 3d shows that lower frequency component ($\omega/\omega_c < 1$, corresponding to the rear part of the pulse) is depleted, while strong RBS peak appears at around half-critical frequency ($\omega/\omega_c \sim 0.5$). No signature of Raman forward scatter (RFS) is observed, as the RFS growth rate is very small on the steep gradient.

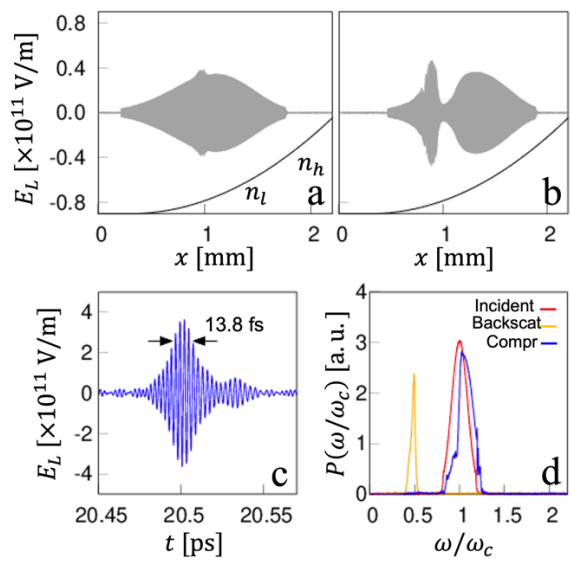

FIG. 3: **a,b**. Depletion of the incident pulse energy in a thermal plasma ($T_e = 10$ eV) by RBS. **c.** Compressed pulse profile in time and **d.** spectra of the incident, the compressed pulse and the RBS signal.

Since the RBS occurs only for $n < n_{1/4}$, cutting out the plasma below the quarter-critical density can remedy energy loss by RBS. In Fig. 4, we examined the intensity ($I$) and the energy loss ratio ($U/U_0$) of the compressed pulse to the incident pulse, varying the incident pulse intensity ($I_0$) and plasma temperature, for full quadratic and quarter-cut plasmas. For the full quadratic plasma, the compression saturates mostly for $I_0 > 10^{14}$ W/cm². With the quarter-cut plasmas, however, the saturation point is considerably delayed to $I_0 > 3 \times 10^{14}$ W/cm². Note that the points marked by '1' and '2' in Fig. 4 correspond to the maximum compressed intensity ($\sim 10^{17}$ W/cm²) and the largest multiplication factor (225 times), respectively, described in Sec. II.

Figure 4b shows the ratio of compressed pulse energy ($U$) to the incident pulse energy ($U_0$). In the incident range $2 \times 10^{14} < I_0 < 8 \times 10^{14}$ W/cm², the ratio is up to twice larger in the quarter-cut plasma compared with the full quadratic plasma. However, the intensity after compression does not scale by the same factor (Fig. 4a). The difference originates from changed path length of photons as a result of the plasma cut. By adjusting the slope or profile of the density, the compression rate can be further optimized. Here we would like to emphasize that the RBS in PIC simulations are overestimated by numerically enhanced thermal noise. Hence it is possible that the saturation of compression or energy in Figs. 3 and 4 is delayed even more in real experiments, i.e. a better situation.

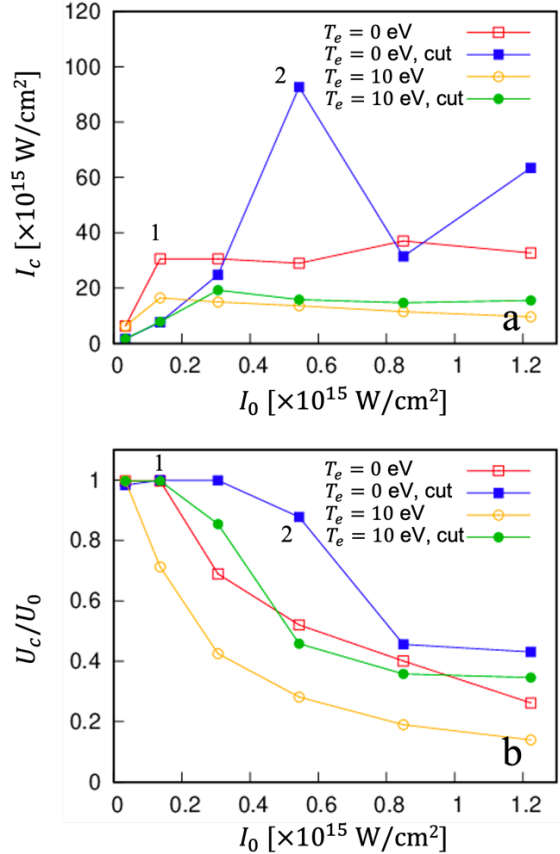

FIG. 4: **a**. Intensity of compressed pulse vs. incident intensity for 0 and 10 eV, and full-quadratic and quarter-cut plasma. **b**. Ratio of compressed pulse energy to the incident pulse energy. $U/U_0 = 1$ is consistent with no energy loss during compression.

## IV. EFFECTS OF FLUCTUATION AND INSTABILITIES

In this section, we theoretically examine the effects of density fluctuation and laser-plasma instabilities, which are potentially harmful to pulse compression.

**Density fluctuation:** Local bumps or troughs in the plasma density profile can unfavourably affect the pulse compression as photons are reflected earlier (by bumps) or later (by troughs) than from unperturbed density profiles. Change in the reflection time results in a spread of photons with respect to their full-compressed position. To examine the spread of photons, we place a cosine-shaped bump around the critical point $l_c$, i.e. $n = n_0 + \delta n_f \cos k_f(x - l_c)$. Here $n_0$ is the unperturbed density (e.g. quadratic) and the second term is non-vanishing over a half wavelength $\lambda_f/2$ around $l_c$. From a ray calculation using the photon's group velocity on an approximately linearly-rising bump, the change of reflection time is

$$\frac{\Delta \tau_c}{\tau} \simeq \frac{\delta n_f}{n_c}, \tag{1}$$

where $\tau$ is the pulse duration. Similar change in the reflection time can be obtained for a density trough. This equation indicates that density fluctuations are detrimental to the efficiency of the compression. As will be examined further, other effects are much less important. Typically, the fluctuation level can be contained to roughly one percent in experiments, and pulse compression by hundred times is achievable, even with fluctuations. For a higher compression rate, a specially designed nonlinear chirp for faster compression and highly tuned experimental techniques for suppressing the fluctuations may be necessary.

Photon spread by fluctuations in the low density region (lower than the critical density) is considerably less significant compared with near-critical fluctuations, as the alternating bumps and troughs tend to average out the deviation of travel-time of photons. From the same group-velocity calculation, but now with $n \ll n_c$, the delay or advance of the photon's travel time is proportional to $\delta n_f \lambda_f/(2cn_c)$. When there are $N \sim L/\langle \lambda_f \rangle$ bumps and troughs randomly distributed, the accumulated change of the travel-time will take on an average displacement as a random-walk, i.e.

$$\Delta \tau_l \sim \sqrt{N} \left\langle \frac{\delta n_f \lambda_f}{2cn_c} \right\rangle \sim \frac{\langle \delta n_f \rangle}{2n_c} \sqrt{\frac{\langle \lambda_f \rangle}{L}} \tau. \tag{2}$$

For a short wavelength fluctuations (i.e. $\langle \lambda_f \rangle/L \ll 1$), $\Delta \tau_l \ll \Delta \tau_c$.

In the above, we used the group-velocity calculation, which does not take into account each photon's bandwidth. However, more rigorous calculation using Maxwell's equations indicates that group-velocity calculation is valid as long as the pulse bandwidth is dominated by the chirp, as in our case.

**Parametric instabilities:** Raman backscatter (RBS) [12] is the fastest instability that can lead to loss of photons. Fortunately, a frequency chirp of the pulse is known to suppress the growth of the RBS [12, 14, 19], because the bandwidth of resonance is very narrow. From the resonant bandwidth $\sim \pi \gamma$ [19], the resonant portion $\Delta \tau_R$ in the pulse duration $\tau$ is approximately given by $\Delta \tau_R/\tau \sim \pi \gamma/\Delta \omega$, where $\Delta \omega$ is the bandwidth of the chirped pulse. The RBS growth rate, $\gamma = a_0 \sqrt{\omega \omega_p/2}$ ($a_0 = eE_L/mc\omega$, i.e. normalized vector potential amplitude of the incident laser field), reaches a maximum at the quarter-critical density ($\omega_p = \omega/2$), leading to $\Delta \tau_R/\tau \simeq \pi a_0 \omega/(2\Delta \omega)$. In the worst case, where the resonant portion of the pulse is completely depleted by RBS, the photon loss is under 10 %, with $\Delta\omega/\omega \sim 0.1$, and we need $a_0 < 0.01$. The energy loss via RBS by 16 %, demonstrated in Fig. 3, which coincides well with this estimation. However, we note that such a limitation in $a_0$ can be removed in diverse ways, e.g., energy loss can be reduced considerably by cutting out the plasma below the quarter-critical density, as we have shown.

Parametric decay into two plasmons [31] has a fast growth rate and narrow growth bandwidth that are comparable to that of RBS. A similar loss of photons (about 10 %) is expected, but as $2\omega_p$ grows at the quarter-critical density, the quarter-cut plasma used for suppression of RBS can also be useful in suppression of two plasmon decay.

Raman forward scattering (RFS) has a growth rate an order of magnitudes lower than RBS, and furthermore, it is strongly suppressed in non-uniform plasma. RFS is not observed in our simulations. Brillouin scattering is a much slower process than the Raman, and is therefore negligible.

**Transverse instabilities:** A wide laser pulse in a high-density plasma can be subject to self-focusing [28], filamentation [26], and Rayleigh-Taylor-like instabilities [29]. The moderate peak amplitude ($a_0 \sim 0.01$) and the supposed wide spot ($r_s \sim 1$ cm) sets the pulse power before compression to tens of terawatt, which is well above the critical power for relativistic self-focusing, $P_c = 17.4\lambda_p^2/\lambda^2$ [GW] [28]. However, for these parameters, the focal length induced by relativistic self-focusing is $z_{focus} \sim \omega r_s/(\omega_p a_0) \sim 100$ cm, which is orders of magnitude longer than the plasma used for the pulse compression. Therefore, relativistic self-focusing can be neglected. Ponderomotive multi-filamentation is another deleterious instability. Its growth rate is given by $\gamma \sim \omega_{pi} a_0$ [26] ($\omega_{pi}$ is the ion plasma frequency). Time for one e-folding of filamentation for $a_0 \sim 0.01$ is ~10 ps or longer for heavier atoms, which is comparable or longer than the pulse duration considered. Therefore, ponderomotive filamentation may be almost negligible or at most marginally affect the compression process. As will be shown, collisional heating is not effective, therefore thermal filamentation [27] should be negligible. The Rayleigh-Taylor-like instability occurs when a plasma is pushed forward by a strong ponderomotive front of the pulse, which is not relevant to our case.

**Collision and inverse-bremsstrahlung:** In a fully ionized plasma, electron-ion collision and subsequent inverse bremsstrahlung can deplete the pulse energy. The number of collisions experienced by a photon traveling toward the reflection point is $N_{col} = \int_0^{\tau_r} \nu_{ei}(t)dt \simeq \nu_{ei}\tau_r$, where $\tau_r$ is the travel time to the reflection point. Spitzer resistivity [30] gives the collision rate $\nu_{ei} \simeq 5 \times 10^{-12} n T_e^{-1.5}$ [eV]. With $n = n_c$, the ratio of $N_{col}$ to the number of oscillations during the same period is $N_{col}/N_{osc} \simeq 5\pi \times 10^{-12} n_c/(\omega T_e^{1.5})$. With $a_0 \sim 0.01$ and $T_e \sim \left(\sqrt{1+a_0^2}-1\right) \sim 20$ eV, we have $N_{col}/N_{osc} \lesssim 0.1$, implying that the pulse can lose at most 10 percent of its energy via collision and inverse-bremsstrahlung on the path toward the reflection point. On the return path, the pulse amplitude is already increased by compression, which significantly reduces the collision rate.

## V. ALTERNATIVE WAYS

We have studied compression of a linearly chirped pulse using parabolic plasma density profiles. However, a free expanding plasma takes on an exponential density profile [32] unless the expansion process is controlled in particular ways. For this case, the pulse chirp can be made nonlinear to obtain the maximum compression, using ultrafast pulse shaping techniques [33]. In experiments, the plasma density profile is not readily controllable, but it is relatively easy to generate an arbitrary pulse chirp to match a non-quadratic density profile.

## VI. CONCLUSION

We have proposed a new and simple method of laser pulse compression using a density gradient plasma. By tailoring the plasma density, the reflection path of high frequency components can be made longer than the path of low frequency components in a frequency-chirped long laser pulse. This idea has been verified using 1-D PIC simulations, showing that a ps laser pulse can be compressed to a fs pulse with a very high efficiency of 99.2 %. The result indicates that a small plasma volume, with only 10 cm in diameter, is sufficient to handle extremely high powers up to 7.3 EW. The simulations show that compression is not affected significantly by thermal noise of the plasma, and deleterious Raman backscattering. Theoretical estimates verify that compression by hundred times can be obtained consistently even with density fluctuations. For the given laser pulse and plasma parameters, the compression process is well outside the deleterious effects of various transverse instabilities such as self-focusing and filamentation. , implying that the compressed beam can preserve the high quality in transverse phase. The proposed idea should pave a new path toward compact exawatt-and-above high power lasers in the future.

**Acknowledgments**

This work was supported by the Basic Science Research Program through the National Research Foundation (NRF) of Korea (Grant number NRF-2016R1A5A1013277, NRF-2020R1A2C1102236, NRF-2022R1A2C3013359). DAJ, SY, BE acknowledge support from the U.K. EPSRC (grant number EP/N028694/1) and received funding from the European Union's Horizon 2020 research and innovation programme under grant agreement no. 871124 Laserlab-Europe. SRY and BE acknowledge support from the STFC (grant number ST/G008248/1).

## DATA AVAILABILITY

The data that support the findings of this study and used for producing figures are freely available for reproduction from website http://cpl.unist.ac.kr/data/compression.html.